\documentstyle[pra,aps,multicol,epsf]{revtex}
\def\addcontentsline#1#2#3{\relax}
\begin{document}
\draft
\title{\Large\bf Orbital Liquid in Perovskite Transition-Metal Oxides}
\author{Sumio Ishihara, Masanori Yamanaka, and Naoto Nagaosa}
\address{Department of Applied Physics, University of Tokyo,
Bunkyo-ku, Tokyo 113, Japan}
\date{\today}
\maketitle
\begin{abstract}
We study the effects of the degeneracy of the $e_g$ orbitals
as well as the double exchange interaction with $t_{2g}$ spins
in perovskite transition-metal oxides. 
In addition to the spin field ${\vec S}_i$, the isospin field
${\vec T}_i$ is introduced to describe the orbital degrees of freedom.
The isospin is 
the quantum dynamical variable,
and is represented by the boson with a constraint.
The dispersion of this boson is flat along $(\pi/a,\pi/a,k_z)$ 
($a$: lattice constant) and the other
two equivalent directions. This enables the orbital disordered phase 
down to low temperatures. We interpret some of the anomalous experiments,
i.e., optical absorption and d.c. resistivity, in the low temperature
ferromagnetic phase of La$_{1-x}$Sr$_x$MnO$_3$ with $x > 0.2$
in terms of this orbital liquid picture. 
\end{abstract}


%\narrowtext
\multicols{2}

The colossal negative magneto-resistance observed in 
transition-metal oxides with perovskite structure,
e.g., the double exchange
system La$_{1-x}$Sr$_x$MnO$_3$, has revived the interest in 
these systems \cite{cha,hel,toku,jin}. 
The valency of Mn ion is Mn$^{3+}$
for $x=0$, whose electron configuration is $(t_{2g})^3(e_g)^1$,
and all the spins are aligned ferromagnetically due to the 
strong Hund coupling.
Because of the strong on-site repulsion, the double occupancy of
$e_g$ orbitals is forbidden \cite{sai,inou},
and the system is a Mott insulator.
When one La is replaced by Sr, one hole is introduced to Mn and Mn$^{3+}$
turns into Mn$^{4+}$. These doped holes, which contribute to the conduction,
can not be described in terms of the one-body theory. 
For a single band case, which is relevant to high-Tc cuprates,
extensive studies have been focused on these doped holes to a Mott insulator.
One of the promising approaches is resonating valence bond (RVB) theory 
\cite{bas,nag}, where the 
spin and charge are separated. 
In the orbital-degenerate case, this idea can be generalized also to 
the orbital degrees of freedom as shown below \cite{Zaanen}.

Experimentally the orbital ordering has been established in the 
low hole-concentration region, i.e., $x \sim 0.0$ \cite{good}, 
where the system is
insulating with the A-type antiferromagnetic long range ordering 
(AFLRO) \cite{wk}.
As $x$ increases the system becomes more and more conductive, and finally
shows the metallic conduction below the ferromagnetic transition 
temperature $T_c$. In this metallic state ($x>0.2$), 
there are several anomalous
features.

\noindent
(1) The optical conductivity shows a narrow coherent peak
up to around $0.02 eV$ followed by a broader ``Drude like" band 
up to around $1 eV$ \cite{opt}. 
The integrated oscillator strength for this two-component
Drude absorption changes down to the very low temperature
where the ferromagnetic moment already saturates, which 
suggests that other degrees of freedom still remain
active. 

\noindent
(2) The photoemission spectra shows only a 
small discontinuity at the Fermi energy $E_F$ 
followed by a gap like behavior \cite{sarm,chai,des}. 
On the other hand, the specific heat is not enhanced with $\gamma \sim 
5 mJ/K^2 mol$ \cite{des}.

\noindent
(3) The anisotropy of the conduction and the spin excitation,
which is expected with the orbital ordering, 
is not observed even at low temperatures \cite{aniso,shir}.

\noindent
(4) There is no symptom of the Jahn-Teller (JT) distortion
in neutron scattering experiments \cite{shir}. 
The displacements of the oxygen ions are independent of the 
temperature across the ferromagnetic transition 
temperature.
\vskip 0.3cm
\noindent
It is noted that the on-site repulsive interaction within each orbital
plays no role in the ferromagnetic state because of the 
Pauli's exclusion principle.
Then it is suggested from the above features that the orbital degrees of 
freedom play some roles in the low temperature phase.

In this paper we study the fluctuation in the 
orbital degrees of freedom in perovskite transition-metal 
oxides.
These degrees of freedom are represented by the isospin $\vec T$,
which is the quantum dynamical variable with 
$T = 1/2$ for the $e_g$ orbitals.
We found that
its fluctuation has a low dimensional character.
Based on this, we propose that the isospin is still disordered, i.e.,
liquid state, in the ferromagnetic state of 
La$_{1-x}$Sr$_x$MnO$_3$ with $x > 0.2$.
We have done some numerical simulations based on this idea, and 
their results are at least encouraging.
\par
In order to study fluctuation in the orbital degrees of freedom, 
we adopt the following Hamiltonian.
\endmulticols
\vspace{-6mm}\noindent\underline{\hspace{87mm}}
\begin{eqnarray}
H=\epsilon_d \sum_{i, \sigma , \gamma} 
    d_{i \gamma \sigma}^\dagger  d_{i \gamma \sigma}^{\phantom{\dagger}}  
   +\sum_{\langle ij \rangle, \sigma , \gamma , \gamma'}  
    \Bigl( t^{\gamma \gamma'}_{ij} 
           d_{i \gamma \sigma}^\dagger  
           d_{j \gamma' \sigma}^{\phantom{\dagger}}  
                      +h.c. \Bigr)
  +H_{e-e} 
   + { 1 \over 2} K\sum_{i , \gamma , \sigma , \sigma'} 
   d_{i \gamma \sigma}^\dagger 
    ({ \vec \sigma})_{\sigma \sigma'}^{\phantom{\dagger}}   
   d_{i \gamma \sigma'}^{\phantom{\dagger}}  
   \cdot {\vec S^{t_{2g}}_i} 
\ .   
\label{eq:ham}
\end{eqnarray}
Here it is assumed that the 2$p$ orbitals of oxygen have been integrated over, 
and only the $d$ orbitals of transition-metal ions are considered.
The operator $d_{i \gamma \sigma}^\dagger $ creates an 
electron with spin $\sigma(=\uparrow, \downarrow)$ 
in the orbital $\gamma(=a,b)$ at site $i$. 
The transfer intensity $t^{\gamma \gamma'}_{ij}$ between $\gamma$ orbital 
in site $i$ and $\gamma'$ orbital in the nearest-neighbor site $j$ 
is calculated by considering the overlap integral between the 
$d$ and $p$ orbitals. 
We choose the up and down isospin state in the orbital space 
as $a = (3z^2-r^2)$ and $b = (x^2-y^2)$ orbitals, respectively. 
Then $t^{\gamma \gamma'}_{ij}$ is explicitly written by 
\begin{equation}
t^{\gamma \gamma'}_{ij}=
t_0
\left( \begin{array} {cc} -{1 \over 4}\ , & {\sqrt{3} \over 4} \\
                       {\sqrt{3} \over 4} \ , & -{3 \over 4} 
       \end{array}  \right)_{\gamma \gamma'} ,  \ \ \  
t_0
\left( \begin{array} {cc} -{1 \over 4} \ , & -{\sqrt{3} \over 4} \\
                       -{\sqrt{3} \over 4} \ , & -{3 \over 4} 
       \end{array}  \right)_{\gamma \gamma'} , \ \ \ 
t_0
\left( \begin{array} {cc} -1 \ , & 0 \\
                       0 \ , & 0 
       \end{array}  \right)_{\gamma \gamma'} ,  \ \
\label{eq:tmat} 
\end{equation}
for ${\vec r_j}={\vec r_i} \pm {\hat x}$, 
${\vec r_j}={\vec r_i} \pm {\hat y}$ and  
${\vec r_j}={\vec r_i} \pm {\hat z}$, respectively. 
The constant $t_0$ is positive, and depends on the distance between the 
$d$ and $p$ orbitals.
The electron-electron interaction 
$H_{e-e}$ is explicitly written by 
\begin{eqnarray}
H_{e-e}=U\sum_{i \gamma} 
        n_{i \gamma \uparrow} n_{i \gamma \downarrow} 
       +U'\sum_{i}  n_{i a} n_{i b}  
       +I \sum_{i,  \sigma , \sigma'} 
         d_{i a \sigma}^\dagger  d_{i b \sigma'}^\dagger
         d_{i a \sigma'}^{\phantom{\dagger}}   
         d_{i b \sigma }^{\phantom{\dagger}}    \ , 
\label{eq:hee}
\end{eqnarray}
\noindent\hspace{92mm}\underline{\hspace{87mm}}\vspace{-3mm}
\multicols{2}\noindent
where $n_{i \gamma \sigma}
=d^\dagger_{i \gamma \sigma}d_{i \gamma \sigma}^{\phantom{\dagger}}$ 
and $n_{i \gamma} = \sum_{\sigma} n_{i \gamma \sigma}$.
The fourth-term in Eq. (\ref{eq:ham}) is the Hund coupling  
between $e_g$ and $t_{2g}$ spins 
and ${\vec S}^{t_{2g}}_i$ is the $t_{2g}$ spin with 
$S=3/2$.  
The electron-electron interaction
$H_{e-e}$ in Eq. (\ref{eq:hee}) is rewritten by using the spin operator 
$\vec S={1 \over 2} \sum_{\gamma \sigma \sigma'} 
d_{\gamma \sigma}^\dagger 
({\vec \sigma})_{\sigma \sigma'}^{\phantom{\dagger}}   
d_{\gamma \sigma'}^{\phantom{\dagger}}  $, 
and 
the isospin operator 
$\vec T={1 \over 2} \sum_{\gamma \gamma' \sigma} 
d_{\gamma \sigma}^\dagger 
({\vec \sigma})_{\gamma \gamma'}^{\phantom{\dagger}}   
d_{\gamma' \sigma}^{\phantom{\dagger}}  $
for the orbital degrees of freedom.
First notice the identities 
$
{\vec S}^2={ 3 \over 4} n-{3 \over 2} 
\sum_\gamma  n_{\gamma \uparrow} n_{\gamma \downarrow} 
+2{\vec S_{a}} \cdot {\vec S_{b}} \ , 
$
and 
$
{\vec T }^2={ 3 \over 4} n-2 {\vec S_{a}} 
\cdot {\vec S_{b}}
-n_{a} n_{b}
+{ 1 \over 2} \sum_{\gamma} n_{\gamma \uparrow} n_{\gamma \downarrow}$, 
where $n=n_a+n_b$. 
Then $H_{e-e}$ is rewritten as  
$
H_{e-e}=- \sum_{i} 
\Bigl( \alpha^{(s)}  {\vec S_i} ^2+\alpha^{(t)} {\vec T_i} ^2
+ \beta {\vec S_{ia}} \cdot {\vec S_{ib}}    \Bigr) \ , 
$
where
$\alpha^{(s)}={2 \over 3}U+{1 \over 3}U'-{1 \over 6}I$, 
$\alpha^{(t)}=U'-{1 \over 2}I$, and 
$\beta=-{4 \over 3} (U-U'-I)$. 
Because (i) $\beta$ is order of $I$ and is
much smaller than $\alpha^{(s)}$ and $\alpha^{(t)}$ and (ii) 
the large $U'$ forbids the simultaneous occupancy of $a$ and $b$  
orbitals, 
we will neglect the 
${\vec S_{ia}} \cdot {\vec S_{ib}}$ term. 
Since $\alpha^{(s)}$ and $\alpha^{(t)}$ are both positive, 
the electron-electron interaction tends to induce  
the spin and isospin moments. 
By introducing two kinds of 
Stratonovich-Hubbard auxiality fields,  
the partition function is given by
\endmulticols
\vspace{-7mm}\noindent\underline{\hspace{87mm}}
\begin{equation}
Z=\int \prod_i \biggl\{\prod_{\sigma \gamma} 
\Bigl[ D{\bar d_{i \sigma \gamma}}(\tau) \ Dd_{i \sigma \gamma}(\tau) \Bigr]
\ D {\vec \phi^{(s)}_i(\tau)} \ D {\vec \phi^{(t)}_i(\tau)} \biggr\} 
e^{-\int d \tau L} \ , 
\label{eq:zzz}
\end{equation}
with  
\begin{eqnarray}
L&=& 
\sum_{i, \gamma, \sigma} {\bar d_{i \gamma \sigma}}(\partial_\tau+\epsilon_d)
                               d_{i \gamma \sigma}
+\sum_{\langle ij \rangle, \sigma ,\gamma ,\gamma'}  
      \Bigl(  t_{ij}^{\gamma \gamma'}
       {\bar d_{i \gamma \sigma}} d_{j \gamma' \sigma}+ 
      {\it h.c.} \Bigr)
+ K\sum_{i} 
 {\vec S_i}^{\phantom{\dagger}} \cdot {\vec S^{t_{2g}}_i} 
 + L_{t_{2g}}
 \nonumber \\
& &\hspace{45mm} -2\sum_{i} \Bigl(\alpha^{(s)} \ {\vec \phi^{(s)}_i} 
    \cdot {\vec S_i}^{\phantom{\dagger}}
           +\alpha^{(t)} \ {\vec \phi^{(t)}_i} 
    \cdot {\vec T_i}^{\phantom{\dagger}} \Bigr)
+\sum_{i} \Bigl( \alpha^{(s)} {\vec \phi^{(s)\ 2}_i}+\alpha^{(t)} 
{\vec \phi^{(t)\ 2}_i} \Bigr) \ ,  
\label{eq:l1}
\end{eqnarray}
\noindent\hspace{92mm}\underline{\hspace{87mm}}\vspace{-3mm}
\multicols{2}\noindent
where $L_{t_{2g}}$ is the Berry phase term of the $t_{2g}$ spins. 
Eq. (\ref{eq:l1}) describes the $e_g$ electrons moving in the background
of two fluctuating fields 
$2\alpha^{(s)} \vec \phi^{(s)}$ and 
$2\alpha^{(t)} \vec \phi^{(t)}$. 
We introduce at this stage 
the approximation which is appropriate for the strong 
correlated case. In this case the magnitude of the local fields 
$\alpha^{(s)} |\vec \phi^{(s)}|$ and 
$\alpha^{(t)} |\vec \phi^{(t)}|$ are 
much larger than the transfer intensity $t_0$, and the electron is 
forced to be aligned in the directions of 
$\vec \phi^{(s)}$ and $\vec \phi^{(t)}$ at each site. 
Thus, it is convenient to rotate the spin and isospin  
axes in each site, in order that its z-axis coincides with 
$\vec \phi^{(s)}$ and $\vec \phi^{(t)}$. 
This is accomplished by the unitary matrices 
$U^{(s)}_i$ and $U^{(t)}_i$ for the spin and orbital 
spaces, respectively, which transform the 
fermion operator as 
$f_{i \gamma' \sigma'}
=(U^{(s)}_i)_{ \sigma' \sigma}^{\phantom{\dagger}}
\ (U^{(t)}_i)_{ \gamma' \gamma}^{\phantom{\dagger}}
 d_{i \gamma \sigma}^{\phantom{\dagger}}  $. 
In this rotating frame the fields 
$\vec \phi^{(s)}$ and $\vec \phi^{(t)}$ are pointing in the direction of 
+z, and accordingly the density of states for 
$f$-fermion are divided into four bands separated by the energy gaps.
When the concentration of the $e_g$ electrons is one per each
transition-metal ion, only the lowest band,
which corresponds to 
$\sigma=\uparrow$ and $\gamma=a$, is occupied and the 
system becomes a Mott insulator.
We keep only this lowest band, because the holes are doped  to it and 
only this is important when the low energy excitations
are concerned. We introduce the spinless and orbital-less hole operator
$h_i$ as $h_i = f^\dagger_{i a \uparrow}$.
The virtual transition processes to the higher bands cause the 
exchange interaction terms $L_J$, i.e., so called $J$ term in the 
$t$-$J$ type models \cite{ishi}. 
Then the effective Lagrangian up to the second-order with respect to the 
electron transfer is obtained as follows, 
\endmulticols
\vspace{-6mm}\noindent\underline{\hspace{87mm}}
\begin{eqnarray}
L&=& \sum_{i} {\bar h_i}(\partial_\tau-\mu_h) h_i
+ \sum_{i} ( 1 - {\bar h_i} h_i ) 
   \biggl( ( U^{(t) \dagger}_i \partial_\tau^{\phantom{\dagger}}
 U^{(t)}_i )_{a a}
        +  (U^{(s) \dagger}_i \partial_\tau^{\phantom{\dagger}}
 U^{(s)}_i )
_{\uparrow \uparrow} \biggr)
\nonumber \\
& & \hspace{80mm}  + \sum_{i}3 (U^{(s) \dagger}_i
 \partial_\tau^{\phantom{\dagger}} 
     U^{(s)}_i )_{\uparrow \uparrow} 
-\sum_{\langle ij \rangle} {\bar h_i} {\widetilde t_{ij}}  h_j
 +L_{J} \ , 
\label{eq:l2}
\end{eqnarray}
\noindent\hspace{92mm}\underline{\hspace{87mm}}\vspace{-3mm}
\multicols{2}\noindent
where the chemical potential $\mu_h$ is determined by the condition
$ \langle h^\dagger_ih_i \rangle = x$.
The term
$3(U^{(s) \dagger}_i \partial_\tau^{\phantom{\dagger}}
 U^{(s)}_i)_{\uparrow \uparrow}$ 
comes from $L_{t_{2g}}$ in Eq. (\ref{eq:l1}).
The Berry phase terms of the original electron $d_{i \gamma \sigma}$ 
generates those for 
the rotated fermions $h_i$, spins, and isospins.
Here we introduce the spinor boson
$z^{(s)}_i = ^t[ z^{(s)}_{i \uparrow}, z^{(s)}_{i \downarrow} ]$
to represent the unitary matrix 
$
U^{(s)}_i=
\left( \begin{array} {cc} z^{(s)}_{i \uparrow}\ , 
& - z^{(s)*}_{i \downarrow} \\
 z^{(s)}_{i \downarrow}\ , & z^{(s)*}_{i \uparrow} 
       \end{array}  \right) \ ,  
$
where 
${\vec \phi^{(s)}_i}/|{\vec \phi}^{(s)}_i| 
= \sum_{\alpha \beta} 
z^{(s)*}_{i \alpha} ({\vec \sigma})_{\alpha \beta}^{\phantom{\dagger}}   
z^{(s)}_{i \beta}$ 
and $\sum_{\sigma} |z^{(s)}_{i \sigma}|^2 = 1$.
Correspondingly we introduce $z^{(t)}_i$ for $U^{(t)}_i$.
Then the Berry phase terms for spins are written as 
$ (U^{(s) \dagger}_i \partial_\tau^{\phantom{\dagger}}
 U^{(s)}_i)_{\uparrow \uparrow} 
= \sum_{\sigma} z^{(s)*}_{i \sigma} \partial_{\tau}^{\phantom{\dagger}}
 z^{(s)}_{i \sigma}$,
and its coefficient
should be regarded as the spin quantum number $2S$.
Then in the undoped case ($x=0$), the quantum numbers of the spin and
isospin are $S=2$ and $T=1/2$, respectively.
Then it is expected that the quantum fluctuation is stronger 
for isospin than spin.

Now the original electron operator $d_{i \gamma \sigma}$
with the constraint of no double occupancy is expressed as
$ d_{i \gamma \sigma} = h^{\dagger}_i z^{(t)}_{i \gamma} 
z^{(s)}_{i \sigma} $, 
which is the generalization of the slave-fermion formalism to the
orbital degenerate case. 
Then we call $h_i$, $z^{(t)}_{i \gamma}$, and $z^{(s)}_{i \sigma}$ 
holon, isospinon, and
spinon, respectively. 
In analogy with the spin-charge separation 
in high-Tc cuprates \cite{nag}, one can consider 
the orbital-charge separation in this formalism, which we discuss below.
Then the effective transfer intensity ${\widetilde t_{ij}}$ is given by
\begin{equation}
{\widetilde t_{ij}}=\Bigl (
\sum_\sigma z_{i \sigma}^{(s) \ast} \ z_{j \sigma}^{(s)} \Bigl )
\Bigl(
\sum_{\gamma, \gamma'} z_{i \gamma}^{(t) \ast} \  
t^{\gamma \gamma'}_{ij} \ 
z_{j \gamma'}^{(t)}
\Bigr) \ . 
\label{eq:ttild}
\end{equation}

To study the nature of the orbital fluctuations in the ferromagnetic state, 
we derive the effective action for the isospinon by 
employing the mean-field approximation.
We first consider the kinetic energy 
$-\sum_{\langle ij \rangle} {\bar h_i} {\widetilde t_{ij}}  h_j$
assuming that $xt_0 \gg J$ where 
$J$ is the typical exchange energy and is of the order of 
$t_0^2/\alpha^{(s)}$ or $t_0^2/\alpha^{(t)}$.
Then in the limit of strong correlation, we have the concentration region
where $J/t_0 \ll x \ll 1$, which we are now interested in.
The exchange terms $L_J$ will be discussed later.
We replace the fermion operators by their average value, i.e.,
$\langle {\bar h}_ih_i \rangle=x$, $\langle {\bar h}_ih_j \rangle=-x$, 
and quench the magnetic fluctuation by 
putting $z^{(s)}_{i \sigma} = ^t[ 1, 0]$. 
The constraint  $\sum_{\gamma} |z^{(t)}_{i \gamma}|^2 = 1$ 
is imposed on average 
by introducing the chemical potential $-\lambda$.
Then the effective Lagrangian of the orbital 
excitation $z^{(t)}_{i \gamma}$ is obtained as 
\endmulticols
\vspace{-6mm}\noindent\underline{\hspace{87mm}}
\begin{equation}
L= 
\sum_{\vec k}
x t_0
\left[ \begin{array} {c} z_{a}^{(t)*}(\vec k) \\
z_b^{(t)*}(\vec k) \end{array} \right]
\left[ \begin{array} {cc} 
-{1 \over 2}(c_x+c_y+4c_z), 
& { {\sqrt{3}} \over 2}(-c_x + c_y) \\
 { {\sqrt{3}} \over 2}(-c_x + c_y), 
& - {3 \over 2} (c_x+c_y) 
       \end{array}  \right] \  
\left[ \begin{array} {c} z_a^{(t)}(\vec k) \\
z_b^{(t)}(\vec k) \end{array} \right]
+ \sum_{\gamma, \vec k} \lambda |z_\gamma^{(t)}(\vec k)|^2. 
\label{zterm}
\end{equation}
\noindent\hspace{92mm}\underline{\hspace{87mm}}\vspace{-3mm}
\multicols{2}\noindent
The eigenvalues  for each $\vec k$ are given by
$\epsilon^{(t)}_{\pm} (\vec k)=
\lambda -xt_0 f_{\pm}(\vec k)$
where
$
f_{\pm}(\vec k) = 
 -(c_x +c_y +c_z) \pm [ 
c_x^2+c_y^2 + c_z^2 - c_x c_y - c_y c_z - c_z c_x
]^{1/2}, 
$
with $ c_x = \cos(a k_x)$ etc.
The minimum of this energy are 
given by the flat dispersion along the axis $(\pi/a,\pi/a,k_z)$ 
and the other two equivalent directions.
This situation is quite contrast to the spin case where 
the dispersion relation is proportional to 
$\cos(ak_x) +\cos(ak_y) +\cos(ak_z) $. 
%
%The above feature is originated from the
%structure of $t^{\gamma \gamma'}_{ij}
%$ given in Eq. (\ref{eq:tmat}).
%
The electron in $x^2-y^2$ orbital can not hop
perpendicular to the xy-plane because the overlap with 
the oxygen $p$ orbital is vanishing due to the symmetry.
( Only $(aa)$ component is non-zero in the third matrix in 
Eq. (\ref{eq:tmat}).) 
%Note that this is the characteristics of the $e_g$ orbitals.
%
%For $t_{2g}$ orbitals, 
%$t^{\gamma \gamma'}_{ij} \propto \delta_{\gamma \gamma'}$
%and the isospin behaves similarly to spin.

This flat dispersion leads to an important consequence.
The chemical potential $\lambda$ is determined by 
$ \langle \sum_{\gamma} z^{(t)*}_{i \gamma} z^{(t)}_{i \gamma} \rangle = 
{1 \over N} \sum_{\gamma k} n_B (\epsilon^{(t)}_{\gamma} (\vec k))=1$ 
where $n_B$ is the bose distribution-function. 
Because of this flat dispersion, the 
chemical potential $-\lambda$ is always negative at finite temperature
although it becomes exponentially small below the effective bose
condensation temperature $T^* \sim x t_0$.
Then the orbital long range ordering does not occur 
down to low temperatures.

 One may think that the physical quantities behave similarly 
to the orbital ordered state when the chemical potential is 
very small and $z^{(t)}$ is almost bose condensed.
However this is not the case because there are three-branches 
of the low lying fluctuations, i.e.,
along $(\pi/a,\pi/a,k_z)$ and two equivalent directions.
In order to show it, we generate the random configuration 
of $z^{(t)}$ at very low temperature where the chemical potential is 
already very small.
In this case only the static component of $z^{(t)}$ is important,
which is obtained by the random numbers generated by the 
Gaussian function $\exp[-\epsilon^{(t)}_\pm (\vec k)/T]$. 
Then ${\widetilde t_{ij}}$ is obtained through Eq. (\ref{eq:ttild}), 
and we diagonalize the holon Hamiltonian for 
that configuration of the transfer.
We study the density of states and the optical conductivity $\sigma_h(\omega)$
of the holon system 
by averaging over 50 random samples for the cubic lattice of 
$8 \times 8 \times 8$.
The results for the optical conductivity is 
shown in Fig.\ \ref{fig1}.
If one look at ${\widetilde t_{ij}}$'s, they fluctuate in sign and magnitude
violently, which explains the incoherent nature of $\sigma_h(\omega)$.

\vspace{3mm}
\begin{figure}[hbtp]
\centerline{\epsfile{file=condmat.eps,width=85mm,height=60mm}}
\caption{
\narrowtext
The optical conductivity of the holon $\sigma_h(\omega)$ 
calculated in the $8 \times 8 \times 8$ system. 
The temperature is chosen to be $T=0.1t_0$. }
\label{fig1}
\end{figure}

Now we consider the gauge-field fluctuations.
There are two kinds of U(1) gauge field corresponding to 
$z^{(s)}$ and $z^{(t)}$, respectively.
The spatial components of which are given by the phase of the 
order parameters, i.e., 
$ \langle z^{(r)*}_{i \lambda}  z^{(r)}_{j \lambda'} \rangle =  
| \langle z^{(r)*}_{i \lambda}  z^{(r)}_{j \lambda'} \rangle 
| \exp(i a^{(r)}_{ij})$,
($r = s,t \ \ \lambda, \lambda'=\sigma, \gamma$).
The time components, on the other hand, are the Lagrange multiplier fields
which impose the constraint $\sum_\lambda |z^{(r)}_{i \lambda}|^2 = 1$.
In the ferromagnetic state, $z^{(s)}$ is bose condensed, and 
$a^{(s)}$ becomes massive, while $a^{(t)}$ remains massless in the 
orbital disordered state.
More importantly the Ioffe-Larkin composition rule \cite{ioff} 
applies also the orbital-charge separated state, and the 
optical conductivity $\sigma(\omega)$ is given by
$ \sigma(\omega) = 
\sigma_{h}(\omega)\sigma_t(\omega)/(\sigma_h(\omega)+ \sigma_t(\omega)) $. 
Then the characteristic energy of the orbital fluctuation, which is much 
smaller than that of holons, 
can manifest itself in $\sigma(\omega)$ through 
$\sigma_t(\omega)$.
The photoemission spectrum, on the other hand, is given by the 
convolution of those for holon and isospinon, 
because the Green's function $G$ in space-time representation is 
the product $G(r,\tau) = G_h(r,\tau)G_t(r,\tau)$.
According to the flat dispersion of the isospinon, the holon dispersion is
averaged over and the Fermi edge will be smeared out as observed
experimentally.

 Here we discuss several possible effects which break the 
flat dispersion and/or make the orbitals to be long range ordered.
One possibility is so-called $J$ terms, i.e., 
$L_{J}$ in Eq. (\ref{eq:l2}).
It is noted that the transfer matrix $t_{ij} = ( t^{\gamma \gamma'}_{ij})$
does not commute with any of the Pauli matrices $\sigma^{\alpha}$. 
Then there is no continuous rotational symmetry in the isospin 
space. Neglecting the terms of the order of $x J$, 
$L_J$ in the ferromagnetic phase has the same form as the first term in 
Eq. (\ref{zterm})
with $z^{(t)}_a$ ($z_b^{(t)}$) being replaced by 
$T_z$ ($T_x$) and $xt_0$ by $ -t_0^2/(U'-I)$ \cite{ishi}.
Note that only $T_z$ and $T_x$ appear in $L_J$.
The dispersion has the same form as $z^{(t)}$-boson, i.e., 
$\epsilon_{J \pm}(\vec k)=-{t_0^2 \over U'-I}f_{\pm}(\vec k)$.
Then the minimum of the energy are given by the 
flat dispersion along the axis $(\pi/a,\pi/a,k_z)$ 
and the other two equivalent directions. 
For the lowest dispersion along $(\pi/a,\pi/a,k_z)$, 
$T_z=0$ and only $T_x$ is nonzero, which competes with the fact
that the kinetic energy prefers nonzero $T_z$.
As before the ordering of the isospin is suppressed  
because there are three branches of the flat dispersion with the 
role of $z$'s being replaced by $T$'s.

Applying the mean-field approximation to $L_J$, which gives the 
two-body interactions between the $z^{(t)}$-bosons, 
$L_J$ does not generate a dispersion for the flat branch
because flatness comes from the fact that the hopping of the 
$x^2-y^2$ electron along the z-axis is forbidden. 
We also note that the orbital angular momentum is quenched in 
$e_g$ orbitals, and the spin-orbit interaction is ineffective.
 
The orbital liquid, on the other hand, 
should be very sensitive to the anisotropy between 
x-, y-, and z-axis. In this respect, in the layered materials, e.g., 
La$_{1-x}$Sr$_{1+x}$MnO$_4$ and
La$_{2-2x}$Sr$_{1+2x}$Mn$_2$O$_7$, the orbital liquid 
should be absent. Experimentally the incoherent band 
extending to $\omega \sim 1eV$ seems to be
missing in $\sigma(\omega)$ in these materials \cite{ishik}.
Another test on our scenario is the effects of the uniaxial pressure.
We expect $\sigma(\omega)$ becomes anisotropic and becomes sharp in the
more conductive plane.
\par
The authors would like to thank S. Maekawa,  
Y.~Tokura, J.~Zaanen, T.~Ishikawa and Y.~Endoh for valuable discussions. 
This work was supported by Priority Areas Grants from 
the Ministry of Education, Science and Culture of Japan, 
and the New Energy and Industrial Technology Development 
Organization (NEDO).  
One of the authors (M.Y.) is supported by Research Fellowship
for Young Scientists of the JSPS. 

%
%\vfill
%\eject
%\noindent
%Figure captions
%\par \noindent
%\par 
%\noindent
\end{document}